# Fooling a Real Car with Adversarial Traffic Signs

*Nir Morgulis, Alexander Kreines, Shachar Mendelowitz, Yuval Weisglass*

Harman International, Automotive Security Business Unit

### Abstract

The attacks on the neural-network-based classifiers using adversarial images have gained a lot of attention recently. An adversary can purposely generate an image that is indistinguishable from a "good" image for a human being but is misclassified by the neural networks. The adversarial images do not need to be tuned to a particular architecture of the classifier – an image that fools one network can fool another one with a certain success rate.

The published works mostly concentrate on the use of modified image files for attacks against the classifiers trained on the model databases. Although there exists a general understanding that such attacks can be carried in the real world as well, the works considering the real-world attacks are scarce. Moreover, to the best of our knowledge, there have been no reports on the attacks against real production-grade image classification systems.

In our work we present a robust pipeline for reproducible production of adversarial traffic signs that can fool a wide range of classifiers, both open-source and production-grade in the real world. The efficiency of the attacks was checked both with the neural-network-based classifiers and legacy computer vision systems. Most of the attacks have been performed in the black-box mode: the adversarial signs produced for a particular classifier were used to attack a variety of other classifiers. The efficiency was confirmed in drive-by experiments with a production-grade traffic sign recognition systems of a real car.

## 1 Introduction

Front facing camera modules can be found in almost any modern vehicle, used in variety of tasks such as traffic lane assist [1–3], traffic sign recognition (TSR) [4–7], and object detection [8–11]. As we move on towards autonomous vehicles, the vision systems of the vehicle will need to do more complicated tasks and will take a big part in the decision making and path planning of the vehicle [12,13]. One of the enablers of this technological leap is the rapid progress of Machine Learning algorithms for computer vision [14].

The use of computer vision algorithms based on neural networks introduces a new type of vulnerability into the system. By creating intentional weak perturbations that get strong response in the output of the classifier, an attacker can produce malicious images or objects that can fool such systems [15,16]. Such attacks can be either based on the knowledge of the architecture of the attacked network (white-box approach [17–20]) or be completely architecture-independent (black-box approach [21–24] ). The white-box attacks are understandably stronger than black-box. There are publications that show that detection and segmentation networks are also vulnerable [20,25,26].

In several recent studies, there were attempts to create real-world objects that when perceived through a camera, can fool a trained classifier. Athalye et al. [27] developed the Expectation over Transformation (EOT) algorithm and used it to print 3D objects that can fool a classifier when observed from a variety of viewing angles.

Another notable example of a successful real-world physical attack (Deceiving Autonomous caRs with Toxic Signs, DARTS) can be found in the work done by Sitawarin et al. [28]. They have demonstrated a successful physical adversarial attack on real-sized traffic signs. They created a pipeline for upscaling adversarial perturbation to a printable size, and used the real-size printed signs to fool a classifier that gets a video feed from the viewing angle of a front-facing camera in a real vehicle.

## 1.1 Contribution

Our work shows how to utilize adversarial attacks to attack real-life systems in the physical world. Our research goal was to understand whether commercial TSR systems installed in vehicles could be susceptible to adversarial traffic signs in realistic real-world scenarios. This is the first time that a successful physical adversarial attack on a commercial classification system has been demonstrated.

- We have extended the adversarial creation pipeline proposed by Sitawarin et al. [28] with improved random augmentation techniques and the ability to create perturbations which are tailored to speed limit traffic signs and are therefore less perceptible to a human viewer.
- We have created a pipeline that allows robust production and evaluation of printing-size adversarial signs in the black-box mode.
- We proved that adversarial signs produced by our proposed technique can fool a commercial car perception system in a consistent manner in real-world driving conditions.

# 2 Background

## 2.1 Adversarial Attacks

When attacking a classifier $f: X \rightarrow Y$, a function that maps images from domain $X$ into labels from domain $Y$, the attacker takes an image $x \in X$, and uses it to derive an adversarial image $\tilde{x}$, usually by adding a small (and ideally imperceptible) perturbation to the original image: $\tilde{x} = x + \varepsilon$. When such an attack is successful the adversarial image will produce a different classifier output than the source image that it was created from $f(x) \neq f(\tilde{x})$.

Adversarial attacks can be performed in the white-box setting – where the attacker has full access to the attacked classifier $f$: to its architecture, its parameters ($\theta$) and its outputs. Black-box setting is a scenario where the attacker has no access to the attacked classifier. A black-box attack is naturally a more challenging task for the attacker, who needs to design his attacks in an indirect manner. A common black-box attack technique [22–24] is to rely on the adversarial transferability phenomenon – it was found that adversarial images that can fool a certain classifier in white-box setting, may remain adversarial when targeting a different classifier in black-box setting. An example for a simple and efficient attack method is the Fast Gradient Sign Method (FGSM), proposed by Goodfellow et al. [29]. In the proposed method the attacker computes an adversarial perturbation by calculating the gradient of a loss function $L_f$ w.r.t the pixels of a given image, model parameters $\theta$ and the correct image label $y$. The problem is linearized by performing a single iteration of fixed size determined by a small scalar $\varepsilon_0$ in the sign direction of the gradient:

$$\varepsilon = \varepsilon_0 sign\left(\nabla L_f(\theta, x, y)\right) \tag{1}$$

For targeted attacks, one can simply calculate the loss w.r.t a selected target $\tilde{y}$ and then perform a single iteration towards the desired minima:

$$\varepsilon = -\varepsilon_0 sign\left(\nabla L_f(\theta, x, \tilde{y})\right) \qquad (2)$$

This method requires only a single iteration and therefore it is very efficient. A defender can include it in the training process of his system to improve its robustness against such attacks. This method of defense is called adversarial training. There are several additional variants of the FGSM methods with improved results [30,31].

Following the early work on adversarial images, researches have published their work on defense techniques, looking for ways to harden the protected classifier and improve its robustness to such attacks [32–36]. The proposed defenses showed promising results when defending against most attack techniques.

Carlini and Wagner [17] Have proposed a new attack technique that produced very strong adversarial samples that were able to break all previously published defenses, we refer to their attack as "CW attacks". They have formulated the problem of finding a successful targeted adversarial attack as an optimization problem:

$$\text{Min } d(\tilde{x}, x) \quad S.T \ f(\tilde{x}) = \tilde{y}, \quad \tilde{x} \in H \qquad (3)$$

Where $d$ is a distance function (typically a p norm), $H$ is the constraint on the target space, The classifier $f$ is a highly non-convex function and the optimization problem cannot be solved directly. The authors have proposed to solve an optimization problem with a surrogate loss function $\tilde{L}$ (defined in (4, 5)). $c$ is a constant used to balance the tradeoff between the perceptibility of the attack and its effectivity in fooling the classifier.

$$\text{Min } d(\tilde{x}, x) + c\tilde{L}(\theta, \tilde{x}, \tilde{y}), \qquad S.T \ \tilde{x} \in H \qquad (4)$$

$$\tilde{L}(\theta, \tilde{x}, T) = \max\left(\max\{z(\tilde{x})_i : i \neq \tilde{y}\} - z(\tilde{x})_{\tilde{y}}, -k\right) \qquad (5)$$

Where $z(\tilde{x})_i$ refers to the logit for class $i$, and $k$ is a constant that can be used to control the desired confidence in the target class. It can also be referred to as an "overshoot" parameter.

The goal of the constraint $H$ is to ensure that the optimization output is a valid image $\tilde{x} \in [0,1]^n$ ($n$ is the image dimensions). It can be done by clipping the coordinates after each iteration, or by clipping into the objective function $f(\min(\max(\tilde{x}, 0), 1))$. Instead of simply clipping the output, the authors have chosen to use the following change of variables and optimize with respect to the new variable $\omega$:

$$\varepsilon = \frac{1}{2}(\tanh(\omega) + 1) - x \qquad (6)$$

Since $-1 \leq \tanh(\omega) \leq 1$, than: $0 \leq x + \varepsilon \leq 1$ so $\tilde{x}$ should always have a valid value. This approach smoothens the gradient decent and prevents it from getting stuck in extreme regions.

The CW attack proved to be a very effective approach to produce adversarial images and it is considered to be a very strong attack technique [16]. Carlini and Wagner have demonstrated that using their attack technique as a basis, an attacker can defeat many types of defense methods [37]. It is also shown that the computed perturbations can affect different architectures and therefore can be used in black-box scenarios as well [16,17] with better transfer efficiency than other attack techniques such as FGSM.

## 2.2 Expectation over Transformation (EOT)

The task of performing a successful adversarial attack over real-world classifiers can be challenging. It was found that adversarial images might lose their effectivity when subjected to minor transformations [38]. When an adversarial object is placed in a real-world environment, it can be viewed from variety of angles with different lighting conditions. A successful attack needs to survive all those transformations some of which can be quite aggressive. The challenging nature of this task led some researchers to the conclusion that physical adversarial attacks are not feasible and should not be considered as a threat [39]

To face the challenge of performing adversarial attacks in the physical world Athalye et al. have proposed the Expectation Over Transformations (EOT) [27] method.

The main idea of their approach is that instead of optimizing a perturbation over a single image, the attacker chooses a distribution $T$ of transformation functions $t(x)$, that he wants his adversarial images to be robust against and introduces the transformed images in both the loss function and the distance between the initial and transformed images. He then calculates the mathematical expectation of both loss function and distance. The resulting optimization problem can be written as follows:

$$\mathbb{E}_{t \sim T}[d(t(\tilde{x}), t(x))] < \varepsilon_0, \tag{7}$$

$$\min \mathbb{E}_{t \sim T}[L_f(\theta, t(\tilde{x}), \tilde{y})]$$

The authors demonstrated that with a choice of T distribution which includes realistic space distortions involved in printing out a 2D image of a 3D object, it is possible to produce an adversarial object that can fool a classifier when viewed from a range of viewing angles and distances.

## 2.3 DARTS Pipeline

In [28], the authors built the following pipeline to tackle the challenge of producing printable physical adversarial traffic signs in the real world dimensions:

1. Take a high-resolution image of a traffic sign $x_{HR}$.
2. Generate a mask $M$ which includes only the sign.
3. Re-size both $M$ and $x_{HR}$ to the input size of the attacked classifier.
4. Produce an adversarial image $\tilde{x}$ w.r.t a given target $\tilde{y}$ and resized downscaled $x$, while spending the perturbation budget, calculated as the $p$ norm of the perturbation, only on pixels which are within downscaled $M$.
5. Re-size the perturbation $\varepsilon_{HR} = \tilde{x} - x$ to the original $x_{HR}$ dimensions.
6. The high resolution adversarial image would be $\tilde{x}_{HR} = x_{HR} + \varepsilon_{HR}$

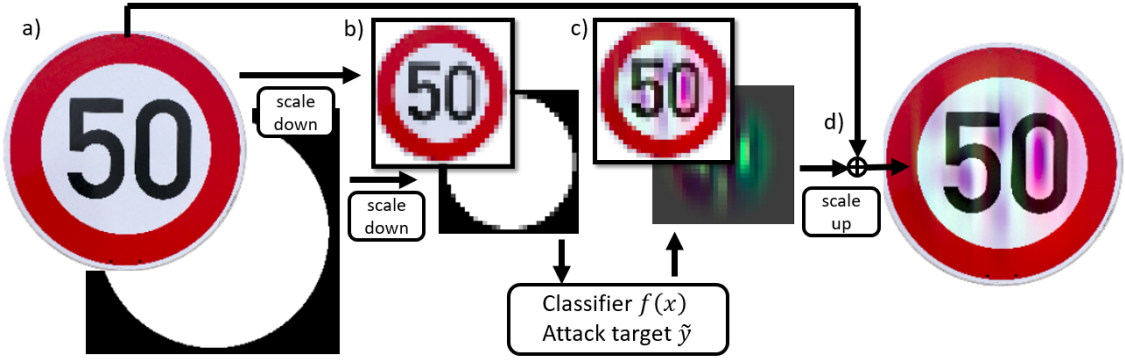

Figure 1: Pipeline for the creation of high resolution adversarial traffic signs with the following stages: a) generate mask, b) resize image and mask into classifier input dimensions, c) generate adversarial image, d) resize the perturbation and add to the original image.

For the production of a robust adversarial image $\tilde{x}$ the authors created a variant of a CW attack combined with EOT:

$$\operatorname*{argmin}_{\varepsilon} \; c \; \frac{1}{B}\sum_{i=0}^{B}[\tilde{L}(\theta, t_i(x + M \cdot \varepsilon), \tilde{y})] + \max(\|\varepsilon\|_p, \varepsilon_{min}) \tag{8}$$

Where $\tilde{L}$ is the surrogate loss function from equation (5) as defined by Carlini and Wagner [17], $B$ is the number of pre-defined transformations in domain $T$ where $t_i \in T$. Transformation space $T$ contains combinations of random image warping and brightness augmentations. The perturbation $\varepsilon$ is applied only to the area defined by the mask $M$, the distance function used is a p norm over $\varepsilon$. The authors added a hyper parameter $\varepsilon_{min}$ to control the minimal perturbations that get penalized. Such a parameter is needed since perturbations which are too small might not survive the transformation to the dimensions of the original, high resolution sign.

## 2.4   Datasets and Classifiers

In order to generate the adversarial images that will later be used for the real-world attack, we first trained two classifiers with Multi-Scale Convolutional Network [7] based architecture, one with the input size of 32x32 pixels, the other with the input size of 128x128 pixels. For training and validating these classifiers, we used the GTSRB [6] dataset which contains more than 50k images of German traffic signs in 43 categories. Our research is focused on the creation of adversarial speed signs since most of the current commercial systems available cannot recognize other types of traffic signs. We used a GTSRB subset which contains only speed signs, in 8 speed categories (11,400 images for training, 1380 for validation, and 4170 for test). We resized all images to uniform dimensions, we had two size variants of the subset – 32X32X3, and 128X128X3.

We were able to achieve 98% accuracy for the 32 pixels classifier, and 98.5% accuracy for the 128 pixels classifier (over the speed signs sub-set).

To verify attack transferability in black-box scenario, we used a classifier with Dense Net [40] based architecture, which was trained on 32 pixel size input. This classier achieves 99.6% accuracy on the speed signs subset.

To create the high quality adversarial speed signs we used an auxiliary data set with 25 high resolution (spanning in the range of 500-2500 pixels) printing templates of traffic speed signs that we have collected/created. All classifiers have 100% accuracy on the auxiliary dataset.

## 3 Expanding the Adversarial Production Pipeline

Our main research question was to understand if a physical adversarial traffic sign can fool a commercial TSR system in a realistic scenario. Sitawarin et al. [28] have showed promising results, and we have decided to use their proposed adversarial production pipeline as the basis for our research. The goal was to increase the chances that the adversarial speed signs produced by the pipeline will work in real-world attacks, and to reduce the perceptibility of the perturbation.

### 3.1 Random Transformations Sampling

The chosen transformations space in the EOT method has a strong effect on the resulting adversarial attack robustness in the real-world conditions. Sitawarin et al. [28] used a random combination of: perspective transformations, brightness adjustments, and image resizing which are predefined before the optimization process begins.

For Brightness transformations, Sitawarin et al. have used a common augmentation practice of adding a single random value to all RGB channels. The main downside of this approach is that such linear transformations are not a good representation of what happens when the lighting conditions are changed in the real world. We wanted to achieve brightness adjustments which better represent the lighting conditions encountered when deploying the adversarial signs outdoors. We converted the images to the YUV color space and used a random value to scale the Y channel which represents brightness. After the augmentation was done, the image was transformed back to RGB for the rest of the pipeline. Scaling the Y channel is a better representation of the physical brightness shift phenomena compared to a general scale of the RGB channels. Another benefit of our approach is that it adds another non-linearity into the transformation space.

One of the hyper parameters in DARTS pipeline is the number of random transformations $B$ which are generated in transformation space $T$. The authors found that as a general rule of thumb, when adding more variations of transformation the resulting adversarial images are more robust. The number $B$ also determines the batch size of the optimization process, so the tradeoff for increasing the amount of transformations is increased computation power and time.

We took a different approach: instead of determining a constant set of random transformations at the beginning of the optimization process, we generate a batch of new random transformations at each iteration.

$$\underset{\varepsilon}{\operatorname{argmin}} \quad c \; \frac{1}{B}\sum_{i=0}^{B}[\tilde{L}(\theta, t_{rand}(x + M \cdot \varepsilon), \tilde{y})] + \max(\|\varepsilon\|_p, \varepsilon_{min}), \; where: \tag{9}$$

$$t_{rand} = T(rand(b_{scale}), rand(m), rand(RS_{scale}))$$

Where transformation space $T$ is now function that generates image transformation functions w.r.t $b_{scale}$ – brightness scaling factor, $m$ – a 3X3 projection transform matrix, and $RS_{scale}$ – a resize scaling factor.

Our approach of continuous randomization of the transformation space effectively increases the number of different transformations used in the image generation. When iterating 3000 steps, the effective number of different transformations would be 3000 * $B$. This allowed us to significantly reduce the value of B; we found that using $B$ values in the range of 32 – 256 gave similar output results. Using small batch size of 32 significantly reduced the computation time needed for robust adversarial sign creation.

### 3.2 Domain Specific Perturbations

When creating an adversarial image, the choice of hyper parameters is usually done as a trade-off between attack effectivity and perturbation perceptibility.

In the optimization problem presented in equations (9), the tradeoff can be controlled with the hyper parameters $c$ (loss function weight), $\varepsilon_{min}$ (minimal perturbation), $k$ (overshoot parameter), and the choice of $p$ norm used to penalize the perturbation. When tuning the hyper parameters, we found that in order to make an adversarial traffic sign that can survive the entire pipeline and successfully fool a classifier after being resized and printed, we need to use relatively aggressive perturbations.

While the attack efficiency can be directly measured by the classification error, measuring the perturbation perceptibility is not straightforward. With a lack of formal definition for perceptibility, the usual approach is to estimate it by measuring the $p$ norm of the perturbation (usually a choice of $L_1$, $L_2$ or $L_\infty$). Such norms do not directly represent the actual perceptibility of the distortion to a human viewer. Drastic shifts in the color space can be immediately spotted, for example: a small red circle on a white background will be prominent to a human viewer compared to a similar circle in a shade of light gray that might be imperceptible for a human. On the other hand, in the RGB color space the red circle (that appears only in a single channel) can have a lower $L_2$ score than a small circle in a shade of gray (that perturbed all 3 channels).

Our work is focused on the speed limit signs, which is a domain with a very limited number of features – typically consists of black digits over a white background. Any perturbation which introduces new types of features into the traffic signs such as new types of colors might immediately appear as out of distribution and perhaps as human intervention.

We added a term to the loss function that encourages perturbations to be in a valid color space for the speed signs domain. We have defined the valid color space as shades of gray. The additional element measures the deviation of the perturbation from the grayscale (grayscale loss term) marked as $L_{gs}$ and is added to the general loss function with a scalar weight $c_{gs}$:

$$\underset{\varepsilon}{\mathrm{argmin}} \ c \ \frac{1}{B} \sum_{i=0}^{B} [\tilde{L}(\theta, t_{rand}(x + M \cdot \varepsilon), \tilde{y})] + \max(\|\varepsilon\|_p, \varepsilon_{min}) + c_{gs} \cdot L_{gs} \qquad (10)$$

Grayscale color space is relatively simple to constrain in the RGB color space. We defined the grayscale loss term as a penalty for differences between the channels of the perturbation, normalized by the image dimensions $H$ and $W$.

$$L_{gs} = \frac{1}{2HW} \sum_{i=1}^{2} \|\varepsilon[:,:,0] - \varepsilon[:,:,i]\|_2 \qquad (11)$$

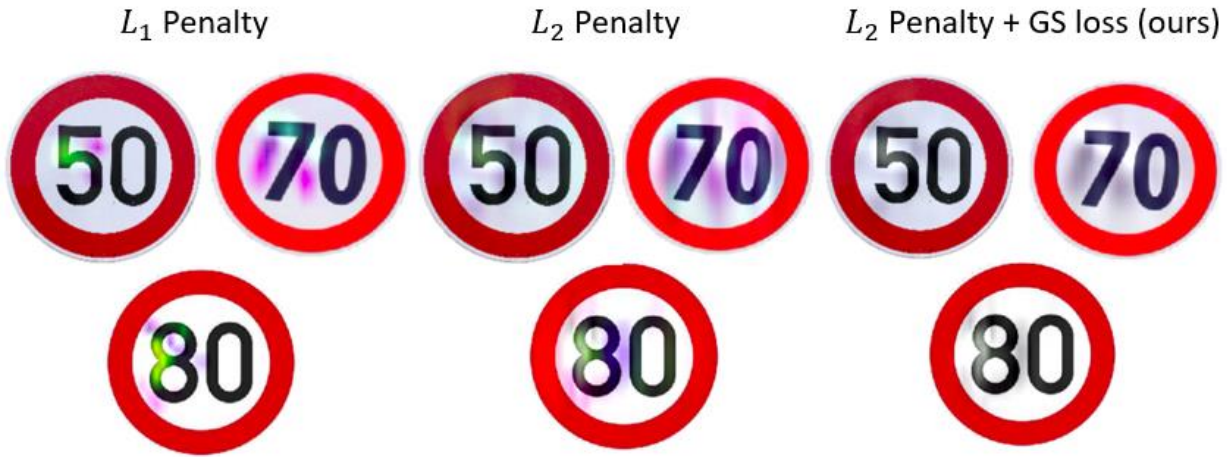

*Figure 2: Examples of the perturbations created when different penalties are used. All signs are targeted to be classified as 30 km/h and were created with equivalent hyper parameters*

In Figure 2 we demonstrate the typical outcomes one might expect when using each of the perturbation penalty policy. It can be seen that our approach produced perturbations that can naturally blend in the image and perhaps be perceived as natural wear or sun damage.

The new hyper parameter $c_{gs}$ controls the strength of the grayscale enforcement. In practice we found that the best results are obtained when the grayscale perturbations are set as hard condition, so we chose a value of $c_{gs} \gg c$ in most of the cases.

### 3.3   TV In the Loop

When creating adversarial images with optimization-based techniques such as CW attacks, the adversarial creation process can be quite slow (1-4 minutes per image), and the optimal hyper parameters can be different at each sample. The creation of a significant amount of spoofed signs even in a white-box scenario can become a slow process requiring human inputs.

Our target was to fool a commercial classifier installed in a vehicle in real-world conditions, meaning that we had no access to the weights or logits of the classifier, and had no knowledge of the classifier architecture; only the final classification output was visible. Our attack approach was to use our own classifier in order to generate the adversarial signs in the white-box mode, and then test the effect of the signs on a different classifier in the black-box mode, expecting a certain yield on account of attack transferability between the classifiers.

Fooling a black-box classifier in real-world conditions as we defined added many uncertainties into the process and thus our expectations on the yield of transferable attacks was very low (our first estimation was <5%). The attack hyper parameters needed to be tuned in a way that maximizes transferability while minimizing the perturbation intensity and perceptibility by a human.

The adversarial production pipeline that was used can indeed create print-sized adversarial signs, but the process is not scalable enough for a successful black-box scenario. The low expected attack transferability yield requires printing a large amount of adversarial traffic signs in order to find the successful signs which can fool the TSR system installed in a car. Tuning the hyper parameters can become a very slow and expensive process. And that is in addition to the logistic challenge of handling potentially hundreds of real-size traffic sign printouts.

We had a practical goal of minimizing the amount of traffic signs that are printed in the attempts to find the successful ones. We added another step into the pipeline in order to better estimate whether a certain attack will survive the transfer into the physical domain.

### 3.3.1 Experimental setup

In order to evaluate the transferability of printed adversarial attacks in black-box scenarios, we added a step where we are simulating the transfer. After tuning the attack hyper parameters over a large batch of traffic signs, the signs were projected on a TV screen. The screen was positioned at angles $\theta$ and $\varphi$ in relation to a camera located in the distance $d$ away from it (as described in Figure 3) and at the same height.

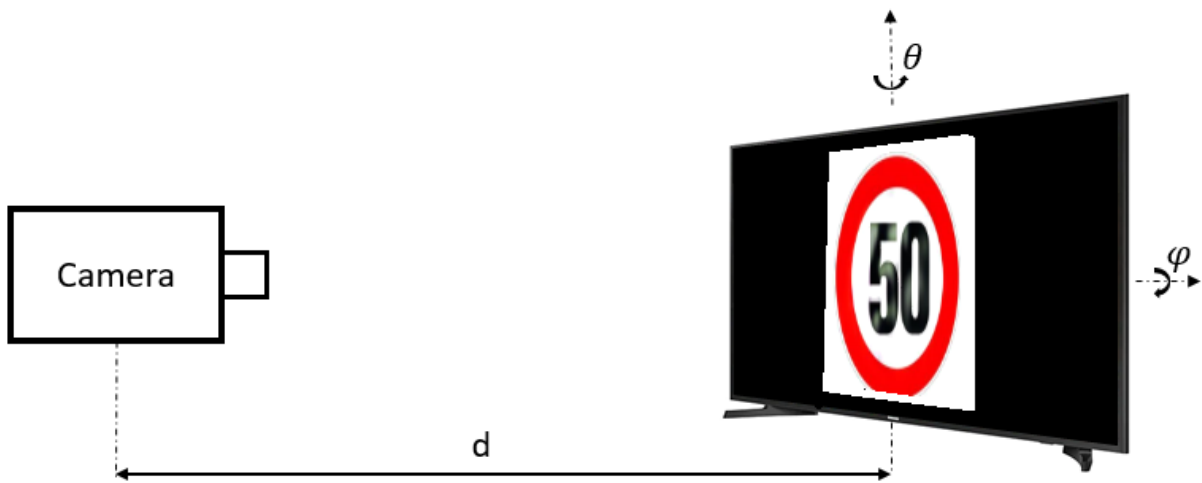

*Figure 3: The schematic representation of the TV in the loop experimental setup*

In order to simulate different imaging conditions, every batch of adversarial signs was taken from a range of positioning combination where: $-45° \leq \theta \leq 45°$, $0° \leq \varphi \leq 15°$ and $7m \leq d \leq 15m$.

At every viewing position, two images were taken for each adversarial traffic sign in the batch: one image of the adversarial sign, and an image of the original traffic sign (clean sign) that was used to produce the adversarial attack.

The resulting images were classified by a classifier that was different from the one used for the image generation thus simulating the black-box attack on the real car TSR. Ideally, we should have used a real production-grade TSR for this purpose but it can only be used when the car is in motion; this is not realistic for the TV-in-the-loop setup.

### 3.3.2 Results

The data collected was used to evaluate the adversarial signs produced after each iteration of tuning the hyper parameter. The physical adversarial success rate (PASR) was evaluated in the following manner (A similar metric was used in [41]):

$$PASR = \frac{\sum_{i=0}^{N}\left(\left(f_{BB}(t_i(\tilde{x})) \neq y\right) \& \left(f_{BB}(t_i(x)) = y\right)\right)}{\sum_{i=0}^{N}\left(f_{BB}(t_i(x)) = y\right)} \quad (12)$$

Where $N$ is the number of viewing angles used to grab the sign images, $t_i$ is the physical transformation of the image as perceived from each viewing angle, and $f_{BB}()$ is a black-box classifier. At some extreme viewing angles the classifier failed to properly classify clean original images of the traffic signs. In order to isolate only the cases of successful adversarial attacks and discard cases where the classifier made an unrelated error, the PASR metric takes into account only viewing angles where the original sign is properly classified. Figure 44 demonstrate a PASR calculation of a single adversarial speed sign.

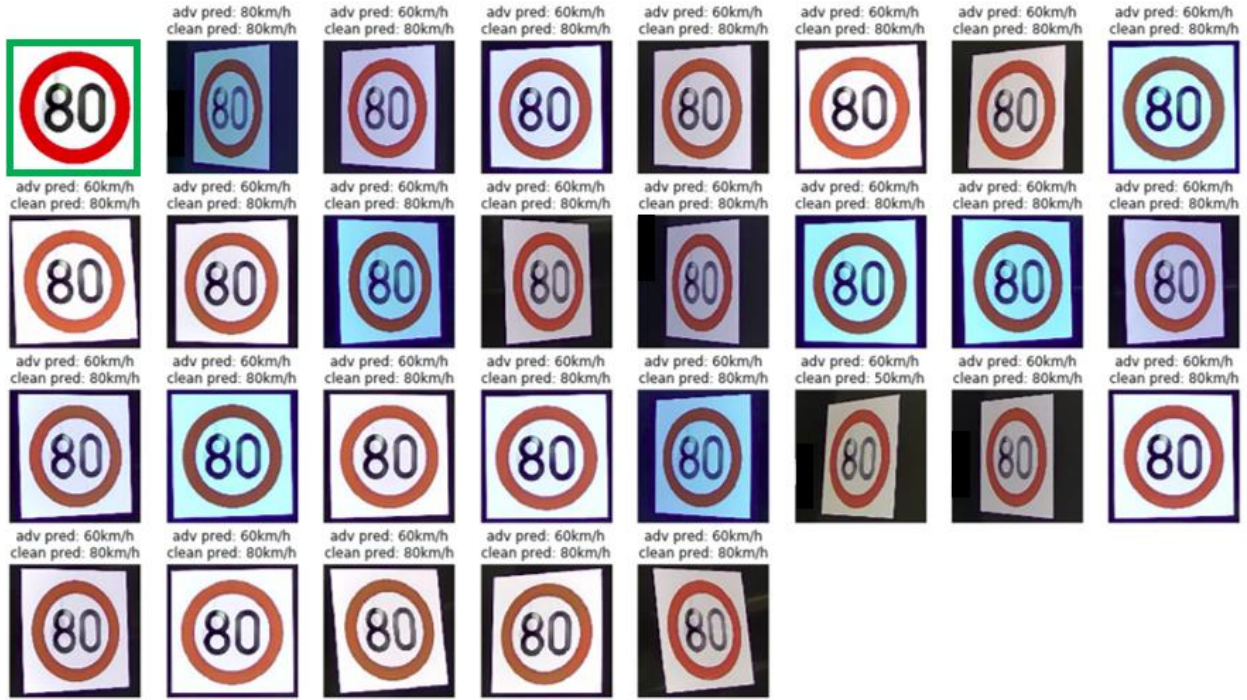

Figure 4: An example of an adversarial sign cropped from images taken at 28 different viewing angles. Title of each image describes the classifier prediction for adversarial and clean versions of the sign. The PASR score of this adversarial sign is 0.96. The original printable version of the traffic sign is on the top left.

The "TV in the loop" method was used to evaluate with a PASR metric a large number of adversarial signs that were produced with variety of hyper parameters. Our assumption was that adversarial attacks which are able to generalize into another black-box classifier after physical augmentations, will have a higher chance of fooling the TSR system installed in the vehicle than adversarial attacks which have failed the tests.

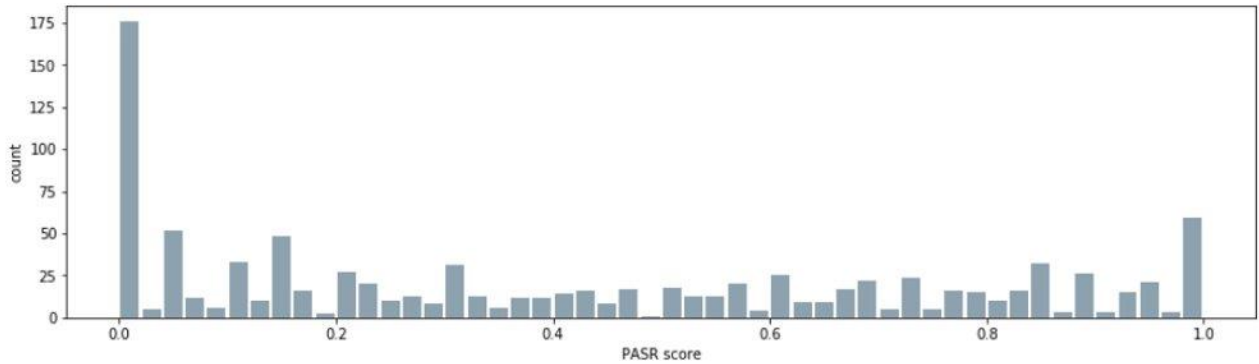

*Figure 5: PASR score distribution across multiple hyper parameter sets.*

Our hyper parameter tuning tactic was to select several combinations of hyper parameters that worked well on selected images, and apply them to all possible combinations of targeted attacks (source 50 km/h with target 100 km/h for example), resulting in a batch of 100-200 adversarial images for each hyper parameter set.

Figure 55 describes the distribution of PASR scores across different sets of hyper parameters. The PASR score was used to filter out the adversarial signs which had low chance for black-box generalization.

The "TV in the loop" approach allowed us to scale up the adversarial production process, and evaluate approx. 1500 adversarial traffic signs that were produced with different hyper parameters (selected in a grid approach) as one batch. We used a threshold of PASR > 0.9 in order to create a subset of approx. 100 adversarial traffic signs which will go to manual evaluation before printing.

We found that there isn't a single set of hyper parameters that can produce the best adversarial images, as each sign that was targeted have reacted differently when attacked. The final subset selected for the field experiment contained adversarial traffic sign images that were created with a variety of hyper parameters.

# 4 Field Experiment

## 4.1 Experimental Setup

The real-world drive-by experiments were performed at the Smart Mobility Analysis and Research Test (SMART) Range in the Negev desert near Beer-Sheva. The SMART Range is a non-profit organization whose aim is building and operation of an Automotive Cyber Security center for the smart mobility solutions in Israel. It is founded by the leading automotive industrial companies and related private Israeli companies, Ben Gurion University of the Negev, and the city of Beer Sheva. The bird's eye view of the SMART Range track is presented in Figure 6.

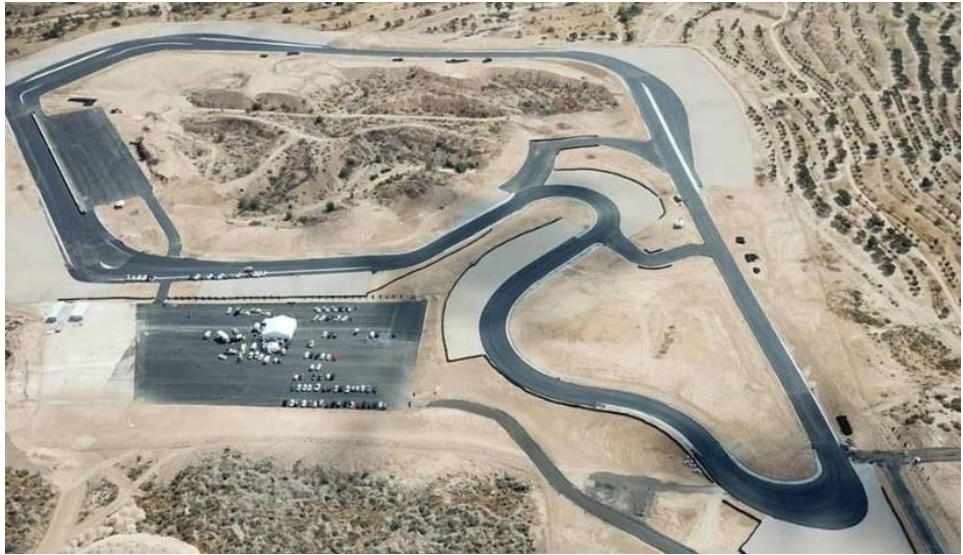

*Figure 7: The bird's eye view on the test field*

The "TV in the loop" pipeline have narrowed down the number of potential successful adversarial signs to a few dozen which have reasonable chance to generalize well in real-world black-box scenario.

We manually narrowed down the list to a final subset of adversarial signs that were printed in real size and positioned around the test field. Our selection was done according the following criteria

- High PASR score
- Perturbation perceptibility – we selected signs with a range of perturbation intensities for our evaluation. We found that the L1 or L2 norms of the perturbations did not reflect the actual attack perceptibility well enough, so a manual evaluation was done as well.

A sample from the final dataset can be seen in Figure 87.

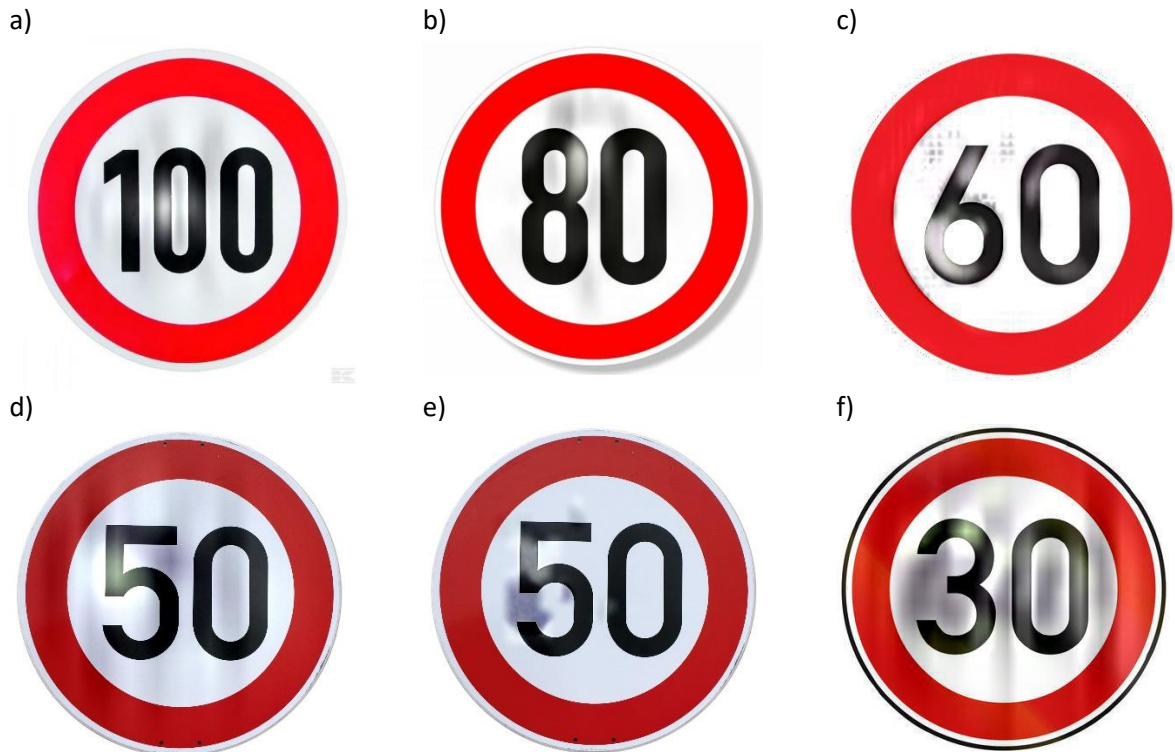

*Figure 8: A sample from the adversarial signs that were tested on the test field. Each sign has its own adversarial target $\tilde{y} =$ : a) 120 km/h, b) 60 km/h, c) 50 km/h, d) 30 km/h, e) 60 km/h, f) 80 km/h*

A combination of spoofed and clean signs were positioned around the track, and were perceived by the TSR system of a car that was driving around the track. The output of the TSR system was recorded from the CANBUS of the vehicle. In addition, since we had no access to the video feed used by the TSR module, we recorded a video of the drive with a dash camera and used it for monitoring and evaluation.

## 4.2 Results

Our experiment was the first time that a commercial TSR module is tested with adversarial traffic signs in such a manner. We have identified that the TSR system may react to an adversarial sign in one of the four following ways:

1. Sign is properly identified with the original source label $y$ (attack failure)
2. Sign is identified with the target label $\tilde{y}$ (attack success)
3. No reaction from the TSR system (attack failure)
4. TSR system freezes for ~1 minute (attack success)

Since only the final output of the TSR system is available, the attack success is evaluated in a binary manner – success or failure. In reaction number 3, we chose to define it as an attack failure because ignoring adversarial traffic signs is actually a valid approach in dealing with such attacks. Reaction number 4 was unexpected, it can be considered as a DOS attack on the TSR module. Without any access to the TSR pipeline we cannot understand the root cause of this behavior, but since it creates a strong disruption of the module operation, we consider this type of reaction as a successful attack.

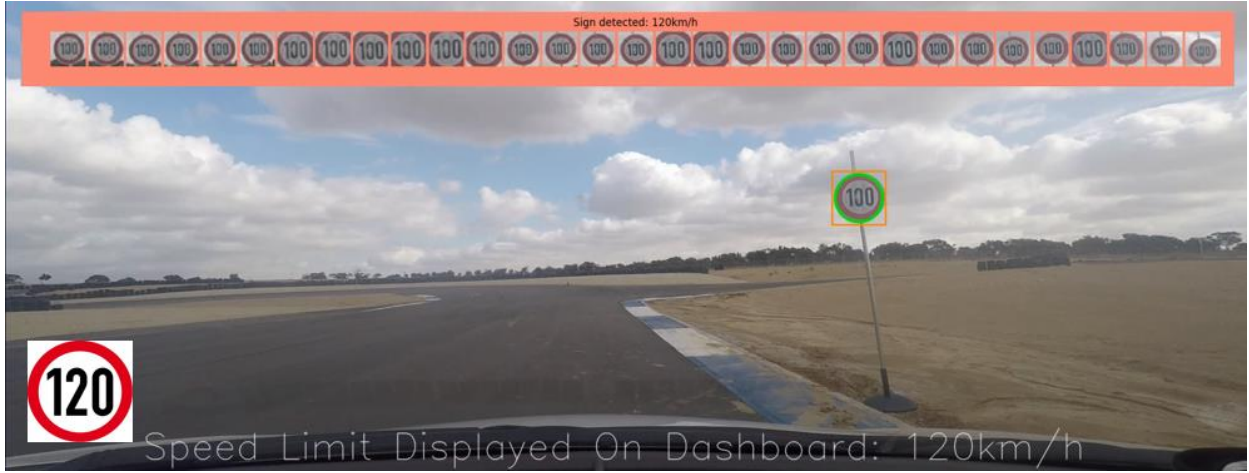

*Figure 9: Adversarial traffic sign on the test track, as seen from a dashboard camera installed on the vehicle*

A combination of adversarial and clean traffic signs were positioned along the track, and we drove the car for several laps at different speeds (range of 30 – 70 km/h), different positions along the lane, and different lighting conditions as the experiment was performed over several hours during the day. A typical frame of the video recorded during the experiments is presented in Figure 9. The banner in the upper part of the frame shows the sequential pictures of the sign recorded by our camera.

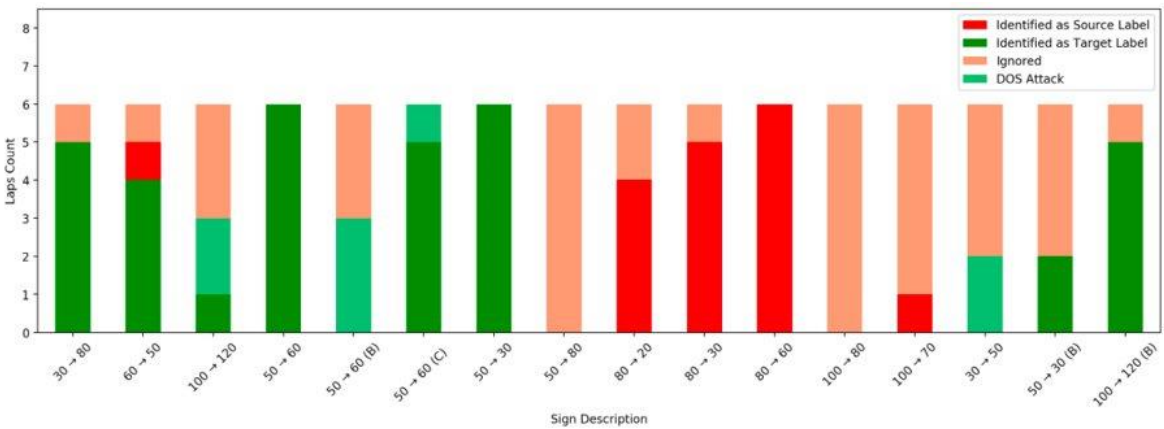

*Figure 10: Summary of the different reactions of the vehicle TSR system when encountered with the adversarial traffic signs on multiple laps.*

Figure 1Figure 10 shows that around 40% of the adversarial signs that have passed our proposed pipeline and were selected to be printed were able to fool the vehicle's TSR system in a complete black-box scenario: no access to classifier weights or architecture, and no knowledge of the classifiers input pipeline. The successful adversarial traffic signs were able to do it in a very consistent manner on the test field and reproduced the TSR behavior in almost every lap.

## 5 Related Work

Sitawarin et al. [28] have proposed a pipeline for the creation of real-sized adversarial traffic signs. The authors have showed that images taken of printed adversarial traffic signs produced by their pipeline can fool a classifier. Our work can be considered as an expansion to the DARTS pipeline.

The authors of [26,41,42] have showed that they can create physical stop signs that after being printed, can successfully fool classification networks in a white-box scenario. Attacks on detection networks (YOLO v2 and Faster RCNN) were demonstrated as well.

Zhang et al. [43] have created a pipeline for creation of adversarial vehicle paint patterns that can fool detection algorithms. The authors have created the attacks and showed their effectivity in a simulation environment. Even though the attacks remained in the simulation domain and were not demonstrated in the real world, the work is still notable. The proposed technique of using a simulator with high quality graphics for the creation realistic attack transformations seems very promising.

A research group from Tencent Keen security lab [44] performed a comprehensive study of reverse engineering for finding security flaws in a Tesla. Among other things, they found weaknesses in the perception systems of the vehicle. They were able to trigger the auto wipers by projecting noise on an electronic display placed in front of the vehicle, thus fooling the visual sensor of the system. They also investigated the lane detection system: it was demonstrated that after application of aggressive blur to a traffic lane the perception system might not detect it, and a fake lane might be produced by placing certain stickers on the road (the latter was not demonstrated yet in real driving conditions).

# 6 Summary

In this study we have presented what to our knowledge is the first time that an adversarial object is used to fool a commercial classification system in a real-world scenario. Our proposed pipeline includes physical transformations as an augmentation method, and domain specific perturbation which reduce the attack visibility. We showed that our pipeline can be used to reproducibly manufacture a significant amount of high resolution adversarial speed signs which transfer to other black-box classifiers in a robust manner. We did an experiment on a test track where we showed that printed adversarial traffic signs produced by our pipeline can fool the TSR system installed on the vehicle in a consistent manner.

# 7 Acknowledgements

The authors are grateful to Dr. Roland Preiss for his interest to the work and highly useful comments and to Etty Haddad and Yoav Etgar for their help in planning, organizing, and implementing the field experiments.

# Appendix Hyper Parameter Tuning

The choice of hyper parameters proved to be a challenging task. We found that the optimal choice of hyper parameters is not always clear, and different images react differently to similar parameter values. In this appendix we describe some of our observations regarding the effects of different hyper parameter choices and the values (or ranges of values) that were used to produce different adversarial traffic signs that were produced in this research.

The attack algorithm used inside our pipeline is based on the CW attack algorithm as described in (9). There are several hyper parameters that need to be selected:

$c$ – The ratio between the loss function and the perturbation penalty. This parameter controls the trade-off between attack strength and the perturbation perceptibility. The usual range that we have used was 2.0 – 15.0.

$B$ – Batch size, which also represents the number of transformations from the transformation domain $T$ used at each iteration. Affects the calculation time and the robustness of the produced adversarial image to distortions when viewed in the real world. We found that our proposed technique of constant randomization produced similar robustness qualities in a range of values, so we chose to use a constant batch size of 32 on all of our tests.

$k$ – Overshoot parameter (5) defining a trade-off between attack strength and perturbation perceptibility, it is also affected by the choice of the value of parameter $c$. Typical values that we used were in the range of 5.0 – 50.0.

$p$ norm – The choice of norm penalty controls the nature of produced perturbation. The choice of $p = 1$ norm usually results in sparse but strong perturbations, while a choice of $p = 2$ will result in smooth perturbation that are spread over larger areas. We have decided to use $p = 2$ in our tests since we found that, when combined with the promotion of grayscale perturbations, it can produce perturbations which can be perceived as natural decay.

$\varepsilon_{min}$ – Minimal perturbation. Perturbation smaller than this value will not get penalized. This parameter is needed, because very small perturbations might disappear when resized to the high resolution dimensions of the original image. We typically chose values in the range of 1.0 - 5.0 when using L2 penalty function ($p = 2$).

$c_{gs}$ – In (10) we introduce a new term of the loss function, which encourages the perturbations to be in shades of gray (the color distribution of the speed signs domain). We found that we get the best result when we set it to very large values ($c_{gs} \gg c$), practically setting the grayscale perturbation color space as a hard constrain.

Image resolution – we attacked two classifiers, trained with different input resolutions: 32 and 128 pixels. The idea was that perturbations produced on a higher resolution image would have better chance to scale up to the full size of the original image. We did not observe a strong evidence that adversarial images produced in higher resolution generalized better with the black-box commercial classifier. Our final printed set had adversarial traffic signs produced in both methods. However it is worth noting that when the 128 pixels classifier was used as a black-box to test transferability of adversarial signs produced by the 32 pixel classifier, it was robust to almost 100% of the attacks. When the reverse test was done, we found that attacks on 128 pixels classifier transferred well and were able to fool the 32 pixels classifier.